\documentclass[a4paper, 10pt, conference]{ieeeconf}      

\IEEEoverridecommandlockouts                              
\usepackage{graphics} 
\usepackage{epsfig} 
\usepackage{rotating}

\title{\LARGE \bf
Does Logarithm Transformation of Microarray Data
Affect Ranking Order of Differentially Expressed Genes?
}

\author{Wentian Li, Young Ju Suh, Jingshan Zhang 
\thanks{W. Li is a Research Scientist with the Robert S Boas Center for Genomics and Human Genetics, Feinstein Institute for Medical Research, North Shore LIJ Health System,
	Manhasset, NY 11030, USA
        {\tt\small wli@nslij-genetics.org}}%
\thanks{Y.J. Suh is a Research Professor  of
	The Research Institute of Natural Sciences, Sookmyung Women's University,
	Seoul 140-742, Korea. 
        {\tt\small yjsprite@yahoo.co.kr}}%
\thanks{J. Zhang is a Senior Statistician at
	Forest Research Institute, Jersey City, NJ 07311, USA
        {\tt\small jingshan.zhang@frx.com}}%
}

\begin{document}

\maketitle
\thispagestyle{empty}
\pagestyle{empty}

\begin{abstract}

A common practice in microarray analysis is to transform
the microarray raw data (light intensity) by a logarithmic
transformation, and the justification for this transformation
is to make the distribution more symmetric and Gaussian-like. Since
this transformation is not universally practiced in
all microarray analysis,  we examined whether the 
discrepancy of this treatment of raw data affect the
``high level" analysis result. In particular, whether
the differentially expressed genes as obtained by 
$t$-test, regularized $t$-test, or logistic regression have altered rank orders
due to presence or absence of the transformation.
We show that as much as 20\%--40\% of significant genes
are ``discordant" (significant only in one form of the
data and not in both), depending on the test being used and the
threshold value for claiming significance. The
$t$-test is more likely to be affected by logarithmic
transformation than logistic regression, and regularized $t$-test
more affected than $t$-test. On the other hand, 
the very top ranking genes (e.g. up to top 20--50 genes, 
depending on the test) are not affected by
the logarithmic transformation.

\end{abstract}

\section{INTRODUCTION}

The number of copies of single-stranded messenger-RNA (mRNA)
can be used to infer the amount of protein product produced
by certain gene, and is called the ``expression level".
Ideally, one would like to count the number of copies of
certain mRNA directly. But in microarray chips, the
amount of a specific mRNA is measured indirectly by
the emission of fluorescence light. It is necessary to 
transform the raw data of light intensity obtained by
optical detection to a summarized quantity
that indicates the expression level. Deriving the
expression level from raw data is called the ``low-level"
analysis, and it can be complicated by the details 
of the technology and chip platform \cite{liwong,irizarry}. 
Reaching conclusions such as the determination of differentially
expressed genes using the expression level data is
called the ``high-level" analysis. 

After the expression level is derived from the raw data,
another preprocessing step is commonly practiced: log-transformation.
The standard motivation for the log-transformation is
that the distribution of the derived expression level
is typically asymmetric with long tail at the high
expression end.  Many parametric statistical tests
require variables to follow a Gaussian/normal distribution.
The log-transformation is an attempt to convert
an asymmetric distribution to a symmetric and Gaussian-like
one. Other transformations for the purpose of ``normality"
are also possible \cite{sokal}, such as square-root, Box-Cox
\cite{boxcox}, and arcsine transformations. In microarray
data, transformations were proposed along the 
line of variance stabilization \cite{durbin1,durbin2}

A novel alternative explanation of the use of
log-transformation might be that human perceive
brightness of light as the logarithm of light
energy, similar to our perceiving loudness of sound 
as the logarithm of sound intensity.  In general, 
all human perception of physical stimuli is proportional 
to the logarithm of amount of stimuli, under the 
names of Weber-Fechner's law \cite{weber,fechner} 
and Steven's law \cite{stevens}. For the light-intensity-derived 
expression level, log-transformation can be 
viewed as a way to measure the ``perception
signal" from the data.

From the statistical point of view, logarithm
transformation can take down an outlier with
extreme high value, thus affecting the group mean.
On the other hand, logarithm transformation or
any 1-to-1 transformation  will not shuffle
the relative order of expression values, thus
will not affect a rank-based test result such
as Wilcoxon-Mann-Whitney test \cite{mann}.
For a specific test or statistical model,
the effect of log-transformation on the
result is not clear, even though we know it
has no effect if the test is rank-based, and
has some effects if there are outliers. For
linear classifiers, the violation of Gaussian
distribution affect some methods more (e.g. Fisher's
linear discriminant analysis, perceptron)
but less so on other methods (e.g.,
logistic regression, support vector machine)
\cite{hastie}.

Another note on investigating the effect of
log-transformation is that one can focus either on
the whole list of genes, or only on the
more interesting top ranking genes. For example,
with a log-transformation, the top 1 and 2
differentially expressed genes may be switched
while the rank of all other genes are unchanged.
Even though the effect of log-transformation
on the whole list of genes could be small, the
minor rearrangement of the top ranking genes
can be crucial in designing the subsequent experiments
such as gene validation by real-time PCR.

We will examine the effect of log transformation
on two or three simple methods for selecting differentially
expressed genes on a real microarray dataset.
Log-transformation is just one factor that change
the apparent value of data, there are other
factors as well such as the normalization 
procedure during the ``low-level" analysis,
change of the probe set design, change of the
microarray platform, etc.

   \begin{figure}[t]
      \centering
	\begin{turn}{-90}
      \resizebox{8.0cm}{6.0cm}{ \includegraphics{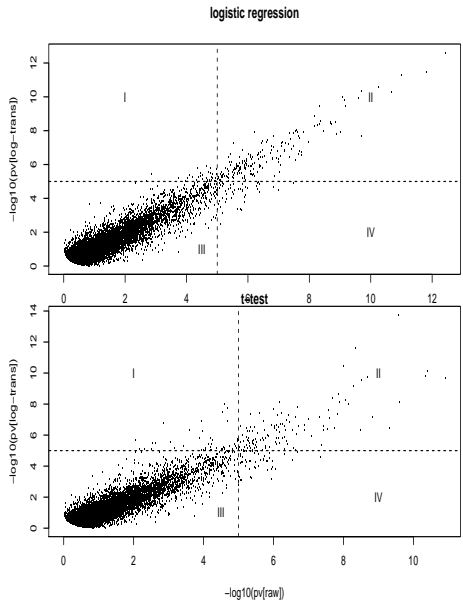} }
	\end{turn}
      \caption{Minus log of $p$-values of tests on log transformed vs. original
data. The $x$ axis is $-\log_{10}(p$-value) for the original
expression data, and $y$ axis is $-\log_{10}(p$-value) for the log-transformed
data. The top plot is for logistic regression and bottom plot
for $t$-test. The four quadrants as split by $x=5$ and $y=5$
are indicated. Each point represents a gene.
	}
      \label{fig1}
   \end{figure}

\section{METHODS AND DATA}

\subsection{Student's $t$-test}

The Student's $t$-test is used here as a representative of
tests that make assumption on variable normality.
We expect the normality requirement is met better 
for the log-transformed data than the original data. The $t$-statistic
is defined as the ratio of the difference of
two group means and the standard error of
this difference: $t= (E_1 - E_2)/\sqrt{ s^2_1/n_1 + s^2_2/n_2}$,
where $E_{1,2}$, $s^2_{1,2}$, $n_{1,2}$ are the
mean, variance, and sample size of group 1 and 2.
The $p$-value given a $t$-statistic value is determined
by the Student's $t$-distribution with degree of
freedom $df$. Usually, $df$ is equal to $n_1+n_2-2$,
but when the variances in two groups are not
equal, a more complicated formula for $df$ can
be used \cite{welsh}. We use such a method as
implemented in the $R$ statistical package ({\sl http://www.r-project.org/}).

\subsection{Logistic regression}

Logistic regression is used to represent statistical 
models which do not have a strong normality requirement.
The advantage for models or tests lacking such a 
requirement is that these are more robust. The 
disadvantage for models without the normality 
requirement is that when the variable is in fact
distributed as Gaussian, these are less ``efficient" 
as classifiers \cite{efron}. The significance of a
single-gene logistic regression can be determined
by a likelihood-ratio test: (-2) log-maximum-likelihood
of the logistic regression model subtract that
of a null model follows a $\chi^2$ distribution
with one degree of freedom, under the null hypothesis.
Thus given the (-2) log-likelihood ratio (called
``deviance"), the $p$-value can be determined using the 
$\chi^2$ distribution.

\subsection{Regularized t-test and significance analysis of microarrays (SAM)}

Since low expression level also leads to low variance,
$t$-statistic can be high due to low expression level. 
Penalized or regularized statistics add an extra 
term $s_0$ to prevent this small variance from inflating the 
statistic: $d= (E_1 - E_2)/(\sqrt{ s^2_1/n_1 + s^2_2/n_2}+s_0)$.
SAM (significance analysis of microarray) is a method
for determining the value of $s_0$ \cite{tusher}. 
SAM test statistic, $d$-score, was calculated by the
SAM package obtained from
{\sl http://www-stat.stanford.edu/~tibs/SAM/}.

\subsection{Microarray data}

The illustrative microarray data is a profiling study of 
rheumatoid arthritis. There are 43 patients
and 48 normal controls, which is more than the 29 patients
and 21 controls used in the previous publication \cite{batli}.
The mRNA was extracted from the peripheral blood mononuclear cells.
The microarray data is obtained from the Affymetrix 
HG-U133A GeneChip with 22,283 genes/probe-sets, and
was normalized by the Affymetrix microarray suite (MAS) program.

\begin{table}
\caption{percentage of discordant genes: (I+IV)/(I+II+IV)}
\label{tab1}
\begin{center}
\begin{tabular}{|c|c|c|c|c|c|c|}
\multicolumn{4}{c}{\em logistic regression} & \multicolumn{3}{c}{\rm t-test} \\
\hline
$p_0$ & I+IV & II & \% (95\%CI) & I+IV & II & \% (95\% CI) \\
\hline
$10^{-9}$ & 0 & 10  & 0\% (0-0) & 7 & 4 & 64\% (35-92) \\
$10^{-8}$ & 6 & 20 & 23 (7-39) & 8 & 11 & 42 (20-64) \\
$10^{-7}$ & 22 & 40 & 35 (24-47) & 21 & 21 & 50 (35-65) \\
$10^{-6}$ & 44 & 84 & 34 (26-43) & 40 & 52 & 43 (33-54) \\
$10^{-5}$ & 82 & 176 & 32 (26-37) & 92 & 119 & 44 (37-50) \\
$10^{-4}$ & 163 & 346 & 32 (28-36) & 170 & 266 & 39 (34-44) \\
0.001  & 328 & 709 & 32 (29-34) & 345 & 593 & 37 (34-40)\\
0.01  & 744 & 1698 & 30 (29-32) & 771 & 1520 & 34 (32-36)\\
\hline
\end{tabular}
\end{center}
\end{table}

\section{RESULTS}

\subsection{Proportion of discordant differentially expressed genes}

Fig.\ref{fig1} shows the minus log of $p$-values of log-transformed
expression data vs that of un-log-transformed (raw)
expression data, for both
logistic regression (top) and $t$-test (bottom). Taking
all genes as a whole, the two sets of $p$-values are highly
correlated (correlation coefficients are 0.94 and 0.93,
respectively).  In order to highlight the 
differences, especially for the high-ranking differentially 
expressed genes, we split the plot into four quadrants 
by a vertical line at $x=a$ and horizontal line at 
$y=a$. The parameter $a=-log_{10}(p_0)$ corresponds 
to gene selection threshold $p_0$ for $p$-values.
For example, the $a=5$ in Fig.\ref{fig1} corresponds
a $p$-value threshold of $p_0=0.00001$.

The genes in quadrants I, II, and IV have at least 
one $p$-value of the two (log and raw data)
smaller than $p_0$, whereas the genes in quadrant II
have both $p$-values smaller than $p_0$. 
If log-transformation has no effect on the gene selection, 
there will be no points in quadrants I and IV. We use the 
percentage of points in I and IV out of all points in I,II, IV 
as a measure of the inconsistency between the test 
results on raw and log-transformed data. If 
points in quadrants I and IV are called ``discordant"
and those in quadrant II ``concordant", this
measure is the percentage of discordant genes among
all differentially expressed genes by either one type
of data.

Table \ref{tab1} shows the discordant percentage and
their 95\% confidence intervals (CI) at various
gene selection threshold $p_0$ (=$10^{-9}, \cdots, 10^{-4}, 0.001, 0.01$).
As expected, the $t$-test result is more affected by the
log transformation than logistic regression: at all $p_0$
threshold values, the percentage of discordant differentially
expressed genes is higher in $t$-test than in logistic
regression. The average discordant percentage at eight
$p_0$ values is 27\% for logistic regression and 44\%
for $t$-test. 

It was however surprising that for logistic regression, 
except for the extremely differentially expressed 
genes (e.g., when $p$-value $< 10^{-9}$, the discordant percentage
is zero), the discordant percentage is not negligible. 
If either one of the raw or log-transformed data is 
used for logistic regression analysis,  as much as 10\%--20\% 
of the claimed differentially expressed genes will not be
claimed so by another data.

   \begin{figure}[t]
      \centering
	\begin{turn}{-90}
      \resizebox{8.0cm}{8.5cm}{ \includegraphics{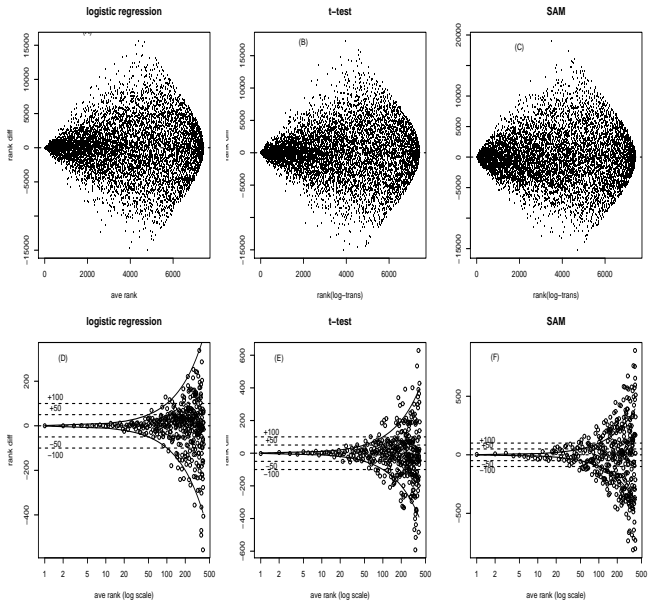} }
	\end{turn}
      \caption{Rank difference $d$ as a function of averaged
rank $R_a$ for all 22283 genes (A,B,C) and for top-400 genes
(D,E,F). Both rank difference $d$ and averaged rank $R_a$ concern
the same gene on two different types of data (raw and log-transformed).
(A) and (D) are results for logistic regression, (B) and (E) are
for $t$-test, (C) and (F) for SAM. The $x$-axis in (D,E,F) is in 
log scale to highlight the top-ranking genes. In (D,E,F), 
$d=50, -50, 100, -100$ and $d=R_a$, $d= -R_a$ lines 
are drawn.
	}
      \label{fig2}
   \end{figure}

\subsection{Ranking change due to log transformation}

The effect of log-transformation can also be examined by
the ranking of a gene in both datasets. If log-transformation
has no effect, the rank of a gene by (e.g.) $p$-value
will be unchanged. We use the notation $R_n(i)$, $R_l(i)$
for the rank of gene-$i$ in the raw and log-transformed data,
and define $R_a(i)$ as the average of the two:
$R_a(i) \equiv (R_n(i)+R_l(i))/2$, and $d(i)$ as the
rank difference: $d(i)= R_n(i)-R_l(i)$. Fig.\ref{fig2} (A,B,C) 
show $d$ vs. $R_a$ for logistic regression, $t$-test, and
SAM (genes are ranked by absolute value of the $d$-score) 
for all 22283 genes.

Fig.\ref{fig2} (A, B,C) indicate that for the whole gene set
there is a similar pattern for all three test-statistics: 
for high- and low-ranking genes, they are high and low ranked in
both raw and log-transformed data (thus smaller rank differences).
As the majority of genes are not differentially expressed,
the overall scattering pattern in Fig.\ref{fig2} (A,B,C) 
may not be as interesting as the behavior near the high-ranking
differentially expressed genes.

To focus on the top-ranking genes, Fig.\ref{fig2} (D,E,F) 
zoom in for the top-400 genes ($x$-axis is in log scale). 
First, we notice that for the very top genes (e.g. up to 
top-10), the ranking is unchanged or changed very little
by the log transformation in any one of the three tests/models. Second, $t$-test
has reached rank-difference of $d=50$ and $d=100$ sooner
(i.e., at a higher ranking) than logistic regression, reconfirming 
our previous conclusion that $t$-test is more likely to 
be affected by log transformation than logistic regressions. 
Using the $d=R_a$ and $d=-R_a$ envelope, we see that
points are more likely to be outside the envelopes for
$t$-test than the logistic regression.  The third 
observation is that SAM test result is affected
even more by log transformation than $t$-test. In
Fig.\ref{fig2} (F), many points are far outside the
envelope region.

\section{CONCLUSIONS AND FUTURE WORKS}

\subsection{Conclusions}

Using one microarray dataset, we have shown that log transformation
may affect results on selecting differentially expressed genes.
If we call all genes that are significant by tests on either raw or
log-transformed data ``differentially expressed genes", and
those genes that are significant in test of only one of the two
types of data ``discordant", the discordant as a proportion of 
the all (discordant and concordant) differentially expressed genes
can be as high as 27\% for logistic regression and 44\% for
$t$-test. The larger discordant percentage for $t$-test confirms
our general understanding that tests that require variable normality
are more likely to be affected by variable transformation.

\subsection{Future Works}

We plan to extend the results here to other public 
domain microarray datasets and to other tests, models, 
and measures for determining differentially expressed genes.

\section{ACKNOWLEDGMENTS}

We thank Franak Batliwalla for providing the data.



\begin{thebibliography}{99}

\bibitem{liwong}
C. Li, W.H. Wong,
``Model-based analysis of oligonucleotide arrays: Expression index 
computation and outlier detection",
{\it Proc. Nat. Acad. Sci.}, vol 98,  pp.31-36.

\bibitem{irizarry}
R.A. Irizarry, B.M. Bolstad, F. Collin, L.M. Cope, B. Hobbs, T. P. Speed,
``Summaries of Affymetrix GeneChip probe level data",
{\it Nucl. Acids Res. }, vol 31, 2003, e15.

\bibitem{sokal}
R.R. Sokal, F.J. Rohlf,
{\it Biometry}, 3rd edition, W.H. Freeman and Co., New York;
1995.

\bibitem{boxcox}
G.E.P. Box, D.R. Cox ,
``An analysis of transformations",
{\it J. R. Stat. Soc. B}, vol 26, 1964,  pp.211-243.

\bibitem{durbin1}
B.P. Durbin, J.S. Hardin, D.M. Hawkins, D.M. Rocke,
``A variance-stabilizing transformation for gene-expression microarray data",
{\it Bioinformatics}, vol 18(suppl 1), 2002, pp.S105-S110.

\bibitem{durbin2}
B. Durbin, D.M. Rocke, 
``Estimation of transformation parameters for microarray data",
{\it Bioinformatics}, vol 19, 2003, pp.1360-1367.

\bibitem{weber}
E.H. Weber,
{\it De pulsi, resorptione, auditu ert tactu.
Annotationes anatomicae et physiologicae},
C.F. L\"{o}hler, Leipzig; 1834.

\bibitem{fechner}
G.T. Fechner,
{\it Elemente der Psychophsik},
Breitkopf \& H\"{a}rtel, Leipzig; 1860.

\bibitem{stevens}
S.S. Stevens,
``On the psychophysical law",
{\it Psychol. Rev.}, vol 64, 1957, pp.153-181.

\bibitem{mann}
H.B. Mann, D.R. Whitney,
``On a test of whether one of 2 random variables is stochastically 
larger than the other", 
{\it Ann. Math. Stat. }, vol 18, 1947, pp.50-60.

\bibitem{hastie}
T. Hastie, R. Tibshirani, J. Friedman,
{\it The Elements of Statistical Learning},
Springer, New York; 2001.

\bibitem{welsh}
B. L. Welsh,
``The generalization of `Student's' problem
when several different population variances are involved",
{\it Biometrika}, vol 34, 1947, pp.28-35.

\bibitem{efron}
B. Efron,
``The efficiency of logistic regression compared
to normal discriminant analysis",
{\it J. Am. Stat. Asso.}, vol 70, 1975, pp.892-898.

\bibitem{tusher}
V. Tusher, R. Tibshirani, C. Chu, (2001):
``Significance analysis of microarrays applied to the ionizing 
radiation response",
{\it Proc. Natl. Acad. Sci.}, vol 98, 2001, pp.  5116-5121.

\bibitem{batli}
F.M. Batliwalla, E.C. Baechler, X. Xiao, W. Li, S. Balasubramaniuan,  H. Khalili, 
A. Damle, W.A. Ortmann, A. Perrone, A.B. Kantor,  P.S. Gulko, M. Kern, R. Furie, 
T. W.  Behrens, P. K. Gregersen,
``Peripheral blood gene expression profiling in rheumatoid arthritis",
{\it Gene and Immunity}, vol 6, 2005, pp. 388-397. 


\end{thebibliography}
\end{document}